\documentclass[11pt]{article}
\usepackage{amssymb,amsfonts,amsmath,latexsym}
\usepackage{color,float,fancyhdr}
\usepackage{graphicx}
\usepackage[latin1]{inputenc}
\usepackage{pstricks,pst-node,pst-text,pst-3d}

\usepackage{ifpdf}

\usepackage{color}
\definecolor{myred}{rgb}{0.5,0,0}
\definecolor{myblue}{rgb}{0,0,0.75}
\definecolor{mygreen}{rgb}{0,0.5,0}

\ifpdf
   \usepackage[pdftex,
               pdftitle={Is Expected Shortfall a good risk measure in practice? A comparison of standard measures},
               pdfsubject={},
               pdfkeywords={},
               pdfauthor={Susanne Emmer, Marie Kratz, Dirk Tasche},
               pdfstartview=FitR,
               breaklinks=true,
               colorlinks=true,
               citecolor=myred,
               linkcolor=myblue,
               urlcolor=mygreen]{hyperref}
\else
   \usepackage{hyperref}
   \usepackage{rotating}
\fi

\def\eqalign#1{\null\,\vcenter{\openup\jot \m@th
  \ialign{\strut\hfil$\displaystyle{##}$&$
     \displaystyle{{}##}$\hfil \crcr#1\crcr}}\,}

\def\eqalignno#1{\displ@y \tabskip=\centering
  \halign to\displaywidth{\hfil$\@lign\displaystyle{##}$
    \tabskip=0pt &$\@lign\displaystyle{{}##}$
     \hfil\tabskip=\centering
     &\llap{$\@lign##$}\tabskip=0pt\crcr #1\crcr}}

\def\leqalignno#1{\displ@y \tabskip=\centering
  \halign to\displaywidth{\hfil$\@lign\displaystyle{##}$
    \tabskip=0pt &$\@lign\displaystyle{{}##}$
    \hfil\tabskip=\centering &\kern-\displaywith\rlap{$\@lign##$}
    \tabskip=\displaywith\crcr #1\crcr}}

\def\negthickspace{\kern-0.277778em }
\newtheorem{theo}{\sc Theorem}[section]
\newtheorem{prop}{\sc Proposition}[section]
\newtheorem{defi}{\sc Definition}[section]

\newtheorem{lem}{\sc Lemma}[section]

\newtheorem{rk}{\sc Remark} [section]
\newtheorem{ex}{\sc Example}[section]
\newtheorem{propties}{\sc Properties}[section]

\def\R{\mathbb{R}}

\def\E{\mathbb{E}}
\def\P{\mathbb{P}}

\def\b{\beta}

\def\n{\noindent }

\numberwithin{equation}{section}

\title{What is the best risk measure in practice? \\A comparison of standard measures}

\author{Susanne Emmer\thanks{CREAR, ESSEC Business School;\, E-mail: susanne.emmer@yahoo.de\newline
Susanne Emmer has been visiting professor at ESSEC Business School in 2012-2013 and is an associated member of CREAR.}, Marie Kratz\thanks{ESSEC Business School, CREAR risk research center; \, E-mail: kratz@essec.edu }, Dirk Tasche\thanks{E-mail: dirk.tasche@gmx.net\newline
Dirk Tasche currently works at the Prudential Regulation Authority (a
division of the Bank of England). He
is also a visiting professor at Imperial College, London.
The opinions expressed in this paper are those of the authors and do not necessarily reflect views of the Bank of England.}} 

\date{January 30, 2015}

\begin{document}

\maketitle

\begin{abstract}
Expected Shortfall (ES) has been widely accepted as a risk measure that is conceptually superior to Value-at-Risk (VaR).
At the same time, however, it has been criticised for issues relating to backtesting. In particular, ES has been found not
to be elicitable which means that backtesting for ES is less straightforward than, e.g., backtesting for VaR. Expectiles have
been suggested as potentially better alternatives to both ES and VaR. In this paper, we revisit commonly accepted desirable
properties of risk measures like coherence, comonotonic additivity, robustness and elicitability. We check VaR, ES and Expectiles
with regard to whether or not they enjoy these properties, with particular emphasis on Expectiles.
We also consider their impact on capital allocation, an important issue in risk management.
We find that, despite the caveats that apply to the estimation and backtesting of ES, it can be considered a good risk measure. 
As a consequence, there is no sufficient evidence to justify an all-inclusive replacement of ES by Expectiles in applications. 
For backtesting ES, we propose an empirical approach that consists in replacing ES by a set of four quantiles, which should allow to make use of backtesting methods
for VaR.\\
\emph{2000 AMS classification}: 62P05; 91B30 
\end{abstract}

\section{Introduction}
\label{se:intro}

Risk Management is a core competence of financial institutions like banks,  insurance companies, investment funds and others. Techniques for the measurement of risk are clearly central for the process of managing risk. Risk can be measured in terms of probability distributions. However, it is sometimes useful to express risk with one number that can be interpreted as a capital amount. Tools that map loss distributions or random variables to capital amounts are called \emph{risk measures}. The following questions are of crucial importance for financial institutions:
\begin{itemize}
	\item What properties should we expect from a risk measure?
	\item What is a `good' risk measure? 
	\item Does there exist a `best' risk measure? 
\end{itemize}
Much research in economics, finance, and mathematics has been devoted to answer those questions. Cram\'er (1930) was one of the earliest researchers on risk capital, introducing ruin theory (Cram\'er \cite{cramer}).  A major contribution was made by Markowitz (1952, \cite{marko}) with modern portfolio theory. The variance of the Profit and Loss (P\&L) distribution became then the dominating risk measure in finance. But using this risk measure has two important drawbacks. It requires that the risks are random variables with  finite variance. It also implicitly assumes that their distributions are approximately symmetric around the mean since the variance does not distinguish between positive and negative deviations from the mean. Since then, many risk measures have been proposed, of which Value-at-Risk  (VaR) and Expected Shortfall (ES) seem to be the most popular. 

In the seminal work by Artzner et al.~\cite{artzner} desirable properties of risk measures have been formalized in a set of axioms. 
Because Expected Shortfall has the important property of coherence, it has replaced VaR, which does not satisfy this property in all cases, in many institutions 
for risk management and, in particular, for capital allocation (Tasche \cite{tasche3}). The Basel Committee on Banking Supervision also 
recommends replacing VaR by ES in internal market risk models \cite{basel2013fundamental}. 
Recently, a study by Gneiting \cite{gneiting} has pointed out 
that there could be an issue with direct backtesting of Expected Shortfall estimates because Expected Shortfall is not \emph{elicitable}. 
Therefore, with a view on the feasibility of backtesting, in recent studies (Bellini et al.~\cite{bellini} and Ziegel \cite{ziegel}) Expectiles have been suggested as coherent and elicitable alternatives to Expected Shortfall. See also Chen \cite{Chen2013} for a detailed discussion of the issue. 

The aim of this paper is to provide a compendium of some popular risk measures based on probability distributions, in order to discuss and compare their properties. We can then provide answers to the questions raised above and study the impact of the choice of risk measure in terms of risk management and model validation. 
 Several recent review papers (e.g. Embrechts and Hofert~\cite{em:ho},  Embrechts et al.~\cite{em:wang}) discuss also the use of risk measures and some of their properties in the context of regulation. Here, we present a panorama of the mathematical properties of four standard risk measures variance, VaR, ES and Expectile, addressed to both academics and professionals.   

We consider a portfolio of $m$ risky positions, where $L_i, i \in \{1,\ldots, m\}$, represents the loss in the $i$-th position. Then, in the generic one-period loss model, the portfolio-wide loss is given by $L=\sum_{i=1}^{m}L_i.$ In this model losses are positive numbers, whereas gains are negative numbers. We assume that 
the portfolio loss variable $L$ is defined on a probability space $(\Omega, {\cal F}, P)$. 

The paper is organized as follows: 
After the introductory section~\ref{se:intro}, Section~\ref{defprop}  recalls the main definitions and properties of what is expected from a risk measure, like coherence, comonotonic additivity, law invariance, elicitability and robustness, before presenting the three downside risk measures that we want to  evaluate in this study.
In Section~\ref{se:prop}, we compare these risk measures with respect to their properties, starting with an overview. After summing up the most important results about subadditivity of Value-at-Risk, we look at different concepts of robustness, discuss the elicitability of Expected Shortfall and Expectiles, and observe that Expectiles are not comonotonically additive.
Section~\ref{se:impact} deals with capital allocation and diversification benefits, important areas of application for risk measures and 
for risk management.
We recall the definition of risk contributions of risky positions to portfolio-wide risk and show how to compute risk contributions for Expectiles.
Furthermore, we introduce the concept of diversification index for the quantification and comparison of the diversification of portfolios. 
We then present in Section~\ref{se:backtesting} methods for backtesting in general and look in more detail at Expected Shortfall. 
The paper ends in Section~\ref{se:concl} with a discussion of the advantages and disadvantages of the different risk measures and a recommendation 
for the choice of a risk measure in practice.
\paragraph{Notation.} $\mathbf{1}_M$ denotes the indicator function of the set $M$, i.e.\ $\mathbf{1}_M(x) = 1$ if $x \in M$ and $\mathbf{1}_M(x) = 0$ if $x \notin M$.

\section{Risk measures: definition and basic properties}
\label{defprop}
 
Risk and risk measure are terms that have no unique definition and usage.
It would be natural to measure risk in terms of probability distributions. But often it is useful to express risk with one number. 
Mappings from spaces of probability distributions or random variables into the real numbers are called risk measures. 
In this paper, a risk measure is understood as providing a risk assessment in form of a capital amount that serves as some kind of buffer against unexpected future losses\footnote{%
See Rockafellar and Uryasev \cite{Rockafellar&Uryasev} for alternative interpretations of risk measures.}. 

\subsection{Coherence and related properties}

Artzner et al.~\cite{artzner} demonstrate that, given some ``reference instrument'', 
there is a natural way to define a measure of risk by describing how close or far  a position is from acceptance by the regulator. 
In the context of
Artzner et al.\ the set of all risks is the set of all real-valued functions on a probability space $\Omega$, which is assumed to be finite. 
Artzner et al.\ define ``the measure of risk of an unacceptable position {\it once a reference, prudent, investment has been specified} as the minimum extra capital $\ldots$ which, invested in the reference instrument, makes the future value of the modified position become acceptable.'' 
Artzner et al.\ call the investor's future net worth `risk'.
Moreover, they state four axioms which any risk measure used for effective risk regulation and management should satisfy. 
Such risk measures are then said to be coherent.
Coherence bundles certain mathematical properties that are possible criteria for the choice of a risk measure.

\begin{defi}\label{coherence}  A risk measure $\rho$ is called {\bf\emph{coherent}} if it satisfies the following conditions:
\begin{itemize}
\item Homogeneity: $\rho$ is {\bf\emph{homogeneous}} if  for all loss variables
$L$ and $h\geq 0$ it holds that 
\begin{equation*}\label{homogene}
\rho(h\,L)=h\,\rho(L).
\end{equation*}
\item Subadditivity: $\rho$ is {\bf\emph{subadditive}} if 
for all loss variables $L_1$ and $L_2$ it holds that
\begin{equation*}\label{subbaditive}
\rho(L_1+L_2)\ \leq\ \rho(L_1)+\rho(L_2).
\end{equation*}
\item Monotonicity: $\rho$ is {\bf\emph{monotonic}} if 
for all loss variables $L_1$ and $L_2$ it holds that
\begin{equation*}\label{monotone}
L_1 \leq L_2\ \Rightarrow\ \rho(L_1)\leq\rho(L_2).
\end{equation*}
\item Translation invariance: $\rho$ is {\bf\emph{translation invariant}} if 
for all loss variables $L$ and $a \in \R$ it holds that 
\begin{equation*}\label{ta}
\rho(L-a)=\rho(L)-a.
\end{equation*}
\end{itemize}
\end{defi}
Comonotonic additivity is another property of risk measures that is mainly of interest as a complementary property to subadditivity.
\begin{subequations}
\begin{defi}
Two real-valued random variables $L_1$ and $L_2$ are said \emph{comonotonic} if there exist a real-valued random variable $X$ (the common risk factor) and non-descreasing functions $f_1$ and $f_2$ such that
\begin{equation*}
	L_1=f_1(X) \quad\text{and}\quad  L_2=f_2(X).
\end{equation*}
A risk measure $\rho$ is {\bf comonotonically additive} if for any comonotonic
random variables $L_1$ and $L_2$ it holds that
\begin{equation*}\label{ca}
\rho(L_1+L_2)\ =\ \rho(L_1)+\rho(L_2).
\end{equation*}
\end{defi}
\end{subequations}
Comonotonicity may be considered the strongest possible dependence of random variables
(Embrechts et al.~\cite{embrechts2002correlation}). Hence, if a risk measure is both
subadditive and comonotonically additive, then on the one hand it rewards diversification 
(via subadditivity) but on the other hand does not attribute any diversification benefits 
to comonotonic risks (via comonotonic additivity) -- which appears quite intuitive.   
Risk measures that depend only on the distributions of the losses are of special interest
because their values can be estimated from loss observations only (i.e.\ no additional information
like stress scenarios is needed).
\begin{defi}
A risk measure $\rho$ is {\bf\emph{law-invariant}} if
\begin{equation*}
P(L_1\leq \ell) = P(L_2\leq \ell), \ \ell \in \R\ \Rightarrow\ \rho(L_1) = \rho(L_2).
\end{equation*}
\end{defi}

\subsection{Elicitability}

An interesting criterion when estimating a risk measure is elicitability, introduced by Osband \cite{osband85} and Lambert et al.~\cite{lambert}, 
then by Gneiting \cite{gneiting}. We briefly recall its definition, which is linked to the one of scoring function. 
For further details, we refer the reader to the recent review on probabilistic forecasting, including the notion of elicitability, 
by Gneiting and Katzfuss \cite{gneitingK}.
For the definition of elicitability we first introduce the concept of strictly consistent scoring functions.

A scoring function aims at assigning a numerical score to a single-valued point forecast based on the predictive point and realization:
\vspace{-2ex}
\begin{defi}\label{scoring}
A \textbf{scoring function} is a function
\begin{eqnarray*}
s: \,\R\times \R & \rightarrow & [0,\infty), \\
 (x,y) & \rightarrow &s(x,y) 
\end{eqnarray*}
where $x$ and $y$ are the \emph{point forecasts} and \emph{observations} respectively.
\end{defi}
\begin{defi}
Let $\nu$ be a functional on a class of probability measures ${\cal P}$ on $\R$:\\
$\displaystyle 
\begin{array}{lccl}
 \nu: &{\cal P} &\rightarrow & 2^{\R} \; \text{(the power set of $\R$)},\\
 & P & \mapsto & \nu(P)\subset\R.
\end{array}
$\\
A scoring function $s:\R\times\R \to [0,\infty)$ is {\bf\emph{consistent}} for the functional $\nu$ relative to the class ${\cal P}$ if and only if, for all $P \in {\cal P}$, $t \in \nu(P)$ and $x \in \R$, 
\begin{equation*}
\mathrm{E}_P\left[ s(t, L)\right]\  \leq\  \mathrm{E}_P\left[s(x,L)\right],
\end{equation*}
$L$ being the loss random variable defined on $(\Omega, {\cal F}, P)$.\\  
The function $s$ is {\bf\emph{strictly consistent}} if it is consistent and 
$$
\mathrm{E}_P \left[ s(t, L)\right]=\mathrm{E}_P  \left[ s(x,L)  \right] 
\quad \Rightarrow\quad x\in \nu(P)
$$
\end{defi}
\begin{defi}
The functional $\nu$ is {\bf\emph{elicitable}} relative to ${\cal P}$ if and only if there is a scoring function $s$ which is strictly consistent for $\nu$ relative to ${\cal P}$.
\end{defi}
\begin{ex} \label{scoringExples}
Standard examples of scoring functions are the following:
\begin{eqnarray*}
s(x,y) &=& (x -y)^2, \ \text{squared error} \\
s(x,y) &=& (\mathbf{1}_{\{x\geq y\}}-\alpha)(x-y)^2\, \mathrm{sgn}(x-y),\ 0<\tau < 1\ \text{fixed}, \ \text{weighted squared error} \\
s(x,y) &=& |x-y|, \ \text{absolute error} \\
s(x,y) &=& s(x,y)=(\mathbf{1}_{\{x\geq y\}}-\alpha)(x-y),\ 0<\alpha < 1\ \text{fixed}, \ \text{weighted absolute error} 
\end{eqnarray*}
Squared, weighted squared, absolute, and weighted absolute errors are strictly consistent scoring functions:  the mean functional is elicited by the squared error, the expectile by the weighted squared error, the median by the absolute error, and the quantile by the weighted absolute error (see Newey and Powell \cite{neweyP87}).
\end{ex}

Elicitability is a helpful criterion for the determination of optimal point forecasts: the class of (strictly) consistent scoring functions for a functional is identical to the class of functions under which (only) the functional  is an optimal point forecast. 
Hence, if we have found a strictly consistent scoring function for a functional $\nu$, we can determine the optimal forecast $\hat x$ for $\nu(P)$ by 
$$
\hat x=\arg \min\limits_{x}\mathrm{E}_P \left[ s(x,L)\right]
$$
Hence elicitability of a functional of probability distributions may be interpreted as the property that the functional can be estimated by  generalised regression. 
Another property, that makes elicitability an important concept, is that it can be used for comparing the performance of different forecast methods (see Gneiting \cite{gneiting} for a detailed discussion).

\subsection{Conditional Elicitability} 
So far we have only distinguished between elicitable and non-elicitable functionals. 
However, it turns out that some useful risk measures are not elicitable but '2nd order' elicitable in the following sense. 

\begin{defi}[Conditional elicitability]\label{cond-elicit} 
A functional $\nu$ of ${\cal P}$ is called {\bf\emph{conditionally elicitable}} if there exist functionals $\widetilde\gamma$ and $\gamma: \mathcal{D} \rightarrow  2^{\R}$  
with $\mathcal{D} \subset \mathcal{P} \times 2^{\R}$ such that
\begin{itemize}
	\item[(i)] $\widetilde\gamma$ is elicitable relative to ${\cal P}$,
	\item[(ii)] $(P,\widetilde\gamma(P)) \in \mathcal{D}$ for all $P\in \mathcal{P}$,
	\item[(iii)] for all $c\in \widetilde{\gamma}(\cal P)$ the functional $\gamma_c: {\cal P}_c \rightarrow  2^{\R},\,
P \mapsto  \gamma(P,c)\subset\R$ is elicitable relative to ${\cal P}_c = \bigl\{P\in\mathcal{P}: (P,c) \in \mathcal{D}\bigr\}$, and
	\item[(iv)] $\nu(P)=\gamma(P,\widetilde\gamma(P))$ for all $P \in \mathcal{P}$.
\end{itemize}
Sometimes, $c$ and $\gamma(P,c)$ respectively are single-valued. In this case we identify the one-point sets 
$c$ and $\gamma(P,c)$ respectively with their unique elements.
\end{defi}
Conditional elicitability is a helpful concept for the forecasting of some risk measures which are not elicitable. In section~\ref{se:ESelicitable} we will study ES as an
 example of a risk measure whose conditional elicitability provides the possibility to forecast it in two steps. Indeed, due to the elicitability of $\tilde\gamma$ we can first forecast $\tilde\gamma(P)$ and then, in a second step, take this result for $\tilde\gamma(P)$ as fix and forecast $\gamma(P,c)$ due to the elicitability of $\gamma_c$.

With regard to backtesting and forecast comparison, conditional elicitability offers a way of splitting up a forecast method into two component methods and separately backtesting   
and comparing their forecast performances. This reflects an approach often applied in practice where a complex forecast method is decomposed into component methods that are 
separately validated. While this approach is attractive for making complex issues tractable it need not necessarily entail optimal choice of forecast models.
\begin{rk} \label{rm:every}
Every elicitable functional is conditionally elicitable.
\end{rk}

\subsection{Robustness} \label{se:robustness1}

Another important issue when estimating risk measures is robustness. Without robustness (defined in an appropriate sense), 
results may not not meaningful, since then small measurement 
errors in the loss distribution can have a huge impact on the estimate of the risk measure. This is why we investigate robustness in terms of continuity. Since most of the relevant risk measures are not continuous with respect to the weak topology, we need a stronger notion of convergence. Therefore, and due to some scaling properties which are convenient in risk management, it is useful to consider the Wasserstein distance when investigating the robustness of risk measures 
(see e.g. Bellini et al.~\cite{bellini}).

Recall that the \emph{Wasserstein distance} between two probability measures $P$ and $Q$ is defined as follows:
\begin{equation*}\label{eq:wasserstein}
d_W(P,Q)=\inf\{E(|X-Y|):\, X\sim P, \,Y\sim Q\}
\end{equation*}
When we call a risk measure robust with respect to the Wasserstein distance, we mean continuity with respect to the Wasserstein distance in the following sense:
\begin{defi}
Let $P_n,\,n\geq 1,$ and $P$ be probability measures, and $X_n\sim P_n,\, n\geq 1$ and $P\sim X.$ A risk measure $\rho$ is called continuous at $X$ with respect to the Wasserstein distance if 
\begin{equation*}
\lim_{n\to\infty} d_W(X_n,X) = 0 \ \Rightarrow \ \lim_{n\to\infty} |\rho(X_n)-\rho(X)| = 0. 
\end{equation*}
\end{defi}
In Section~\ref{se:DiscussionRobust} below, we discuss the robustness properties of some popular risk measures with regard to the Wasserstein distance.

Cont et al.~\cite{cont} use a  different, potentially more intuitive concept of robustness which takes the estimation procedure into account. 
They investigate robustness as the sensitivity of the risk measure estimate to the addition of a new data point to the data set which is used as basis for estimation. It turns out that for the same risk measure the estimation method can have a significant impact on the sensitivity. 
For instance, the risk measure estimate can react in a completely different way on an additional data point if we fit a parametric model instead of using the empirical loss distribution. Thus, robustness in the sense of Cont et al.\ relates more to sensitivity to outliers in the data sample than to mere measurement errors.
Cont et al.~also show that there is a conflict between the subadditivity and robustness (in the Cont et al.\ sense) of a risk measure.

In contrast to robustness based on continuity with respect to weak topology or Wasserstein distance, the concept of Cont et al.\ allows to distinguish between different degrees of robustness. This concept may make it hard to decide whether or not a risk measure is still reasonably 
risk sensitive or no longer robust with respect to data outliers in the estimation sample. However, in finance and insurance, large values do occur and are not outliers or measurement errors, but facts that are parts of the observed process itself. In particular, in (re)insurance, one could argue that large claims are actually more accurately monitored than small ones, and their values better estimated. Thus the question of robustness in the sense of Cont et al.\ may not be so relevant in this context.
That is why for the purpose of this paper we adopt a notion of robustness based on the Wasserstein distance which focuses on small measurement errors.


\subsection{Popular risk measures}

Variance and standard deviation were historically the dominating risk measures in finance. 
However, in the past 20 years or so, they have often been replaced in practical applications by VaR, which is currently the most popular downside risk measure.
\begin{defi}
The {\bf\emph{Value-at-Risk (VaR)}} at level $\alpha\in(0,1)$  of a loss variable $L$  is defined as the $\alpha$-quantile of the loss distribution:
\begin{equation*}\label{VaR}
\mathrm{VaR}_{\alpha}(L)=q_{\alpha}(L)=\inf\{\ell:\,P(L\leq \ell)\geq\alpha\}.
\end{equation*}
\end{defi}
VaR is sometimes criticized for a number of different reasons. Most important are its lack of the subadditivity property and the fact that it completely ignores the severity of losses in the far tail of the loss distribution. The coherent risk measure Expected Shortfall was introduced to solve these issues.
\begin{defi}(Acerbi and Tasche \cite{ac:ta})
The {\bf\emph{Expected Shortfall (ES)}} at level $\alpha\in(0,1)$ (also called Tail Value-at-Risk or Superquantile) of a loss variable $L$ is defined as 
\begin{equation*}\label{ES}
\begin{split}
\mathrm{ES}_{\alpha}(L) &=\frac{1}{1-\alpha}\int_{\alpha}^1 q_u(L)du\\
		      &=\mathrm{E}[L|L\geq q_{\alpha}(L)]+(\mathrm{E}[L|L\geq q_{\alpha}(L)]-q_{\alpha}(L))\left(\frac{\mathrm{P}[L\geq q_{\alpha}(L)]}{1-\alpha}-1\right).
\end{split}
\end{equation*}
 If $\mathrm{P}[L=q_{\alpha}(L)]=0$ (in particular, if $L$ 
is continuous), $\mathrm{ES}_{\alpha}(L)=\mathrm{E}[L|L\geq q_{\alpha}(L)].$
\end{defi}

\subsection{Expectiles}
ES has been shown not to be elicitable (Gneiting \cite{gneiting}). That is why Expectiles have been suggested as coherent and elicitable alternatives 
(Bellini et al.~\cite{bellini}, Ziegel \cite{ziegel}). 
The following definition characterises Expectiles analogously to the familiar characterisation of expected values as solutions to minimisation problems. 
As such, they generalise expected values.
However, this definition is not the most general because it requires the random variable to be square integrable. Therefore we revise it afterwards.
\begin{defi}\label{Expectile}
For $0<\tau<1$ and 
square
integrable $L$, the {\bf $\tau$-Expectile} $e_{\tau}(L)$ is defined as 
\begin{equation*}\label{Expectileeq}
e_{\tau}(L)=\arg\min\limits_{\ell\in\mathbb{R}}\mathrm{E}[\tau\max(L-\ell,0)^2+(1-\tau)\max(\ell-L,0)^2]
\end{equation*}
\end{defi}
Since Value-at-Risk is not coherent and Expected Shortfall lacks direct elicitability, it is interesting to look for risk measures which are coherent as well as elicitable. Possible candidates are Expectiles which we just defined; 
a more general but less intuitive definition is suggested by the following observation: 

\begin{subequations}
\begin{lem}(Newey and Powell \cite{neweyP87}, or Bellini et al.~\cite{bellini})\label{le:expectile} If $L$ is an integrable random variable then
$e_{\tau}(L)$ is the unique solution $\ell$ of the equation
\begin{equation*}\label{eq:DefExpectile}
\tau\mathrm{E}[\max(L-\ell,0)]=(1-\tau)\mathrm{E}[\max(\ell-L,0)].
\end{equation*}
Consequently, $e_{\tau}(L)$ satifies
\begin{equation*}\label{expectileeq2}
e_{\tau}(L)=\dfrac{\tau\mathrm{E}[L\mathbf{1}_{\{L\geq e_{\tau}(L)\}}]+(1-\tau)\mathrm{E}[L\mathbf{1}_{\{L< e_{\tau}(L)\}}]}
{\tau P[L\geq e_{\tau}(L)]+(1-\tau)P[L< e_{\tau}(L)]}.
\end{equation*}
\end{lem}
\end{subequations}
According to Gneiting (\cite{gneiting}, Theorem~10), Expectiles are elicitable on the space of all integrable random variables.

\begin{prop}(Bellini et al.~\cite{bellini})\label{pr:expectile}
Expectiles have the following properties:
\begin{itemize}
\item[(i)] For $0<\tau <1$, Expectiles are homogeneous and law-invariant. As a consequence,
 expectiles are additive for linearly dependent random variables, {\it i.e.}
\begin{equation*}
	\mathrm{corr}[L_1, \,L_2] = 1 \quad \Rightarrow \quad e_{\tau}(L_1+L_2) = e_{\tau}(L_1) + e_{\tau}(L_2).
\end{equation*}
\item[(ii)] For  $1/2\leq \tau < 1$, Expectiles are subadditive (and hence coherent), whereas, for $1/2\geq\tau>0$, they are superadditive.
\end{itemize}
\end{prop}

Ziegel \cite{ziegel} has recently shown that Expectiles 
are indeed the only law-invariant and coherent elicitable risk measures.

From Lemma~\ref{le:expectile} and Proposition~\ref{pr:expectile}, it looks as if Expectiles were ideal to make good for the deficiencies of VaR and
ES. This is not the case, however, because Expectiles are not comonotonically additive as follows immediately from their so-called Kusuoka representation as
given for instance in Ziegel \cite{ziegel}.
\begin{prop}\label{pr:expectileNot}
For  $1/2 < \tau < 1$ Expectiles are not comonotonically additive.
\end{prop}
\textsc{Proof of proposition~\ref{pr:expectileNot}.} \\
If $e_\tau$ were comonotonically additive 
then by Theorem~3.6 of Tasche \cite{tasche2} it would be a so-called \emph{spectral risk measure}.
But then by Corollary~4.3 of Ziegel \cite{ziegel} it would not be elicitable, 
in contradiction to Proposition~\ref{pr:expectile} (iii). 
\hfill $\Box$

\section{Properties of the standard risk measures}
\label{se:prop}

Although considering different risk measures would give a more complete picture of the riskiness of a portfolio, in practice one often has to choose 
one number, which should be reported as a basis for strategic decisions. 
To help for this choice, let us start with Table~\ref{tableProp}
by giving an overview over the considered risk measures and their properties, before coming back to them with more details.

\begin{samepage}
\begin{table}[bt]
\caption{Properties of standard risk measures} \label{tableProp}
\begin{center}
\begin{tabular}{l||c|c|c|c}
 Property & variance & VaR & ES & $e_{\tau} (\text{for }\tau \ge 1/2)$ \\
 \hline\hline
 Coherence & & & x & x\\ 
 \hline
 Comonotonic additivity & & x & x & \\ 
 \hline
Robustness & & x\footnotemark &  & \\
w.r.t.~weak topology & &&&\\ 
\hline
Robustness & x & x & x & x\\
w.r.t.~Wasserstein distance & & & &\\ 
\hline
Elicitability &  & x &  & x\\ \hline
Conditional & x & x & x & x\\ 
Elicitability & &&&
\end{tabular}
\end{center}
\end{table}
\footnotetext{It can be shown that VaR at level $\alpha$ is robust with respect to the weak topology at $F_0$ if $F_0^{-1}$ is continuous at $\alpha$. 
See e.g.\ Theorem~3.7 of \cite{Huber&Ronchetti}.}
\end{samepage}

\subsection{When is Value-at-Risk subadditive?}

The subadditivity property fails to hold for VaR in general, so VaR is not a coherent measure.
 The lack of subadditivity contradicts the notion that there should be a diversification benefit associated with merging portfolios. As a consequence, 
a decentralization of risk management using VaR is difficult since we cannot be sure that by aggregating VaR numbers for different portfolios or business units we will obtain a bound for the overall risk of the enterprise. Moreover, VaR at level 
$\alpha$ gives no information about the severity of tail losses which occur with a probability less than $1-\alpha$, in contrast 
to ES at the same confidence level.

When looking at aggregated risks $\sum_{i=1}^n L_i$, it is well known (Acerbi and Tasche \cite{ac:ta}) that the risk measure ES is coherent.  In particular it is subadditive.
In contrast, VaR is not subadditive in general. 
Indeed, examples (see e.g. Embrechts et al.~\cite{em:lw}) can be given where it is superadditive, i.e.
$$ 
VaR_\alpha\big(\sum_{i=1}^n L_i \big) > \sum_{i=1}^n VaR_\alpha(L_i).
$$
Whether or not VaR is subadditive depends on the properties of the joint 
loss distribution. 
We will  not provide an exhaustive review of results on conditions for the subadditivity of VaR, but present only three of these results in the remainder of this section, namely three standard cases: 
\begin{itemize}
	\item[(i)] The random variables are independent and identically distributed (iid) as well as positively regularly varying.
	\item[(ii)]  The random variables have an elliptical  distribution.
	\item[(iii)] The random variables have an Archimedean survival dependence structure.
\end{itemize}
For further related results, see e.g. Dan\'ielson et al.~\cite{dan} and Embrechts et al.~\cite{em:lw, embrechts2002correlation,
emnw, em:pr}. 

\textbf{Ad (i).} The following result presents a condition on  the tail behavior of iid random variables for Value-at-Risk to satisfy asymptotic subadditivity.
\begin{prop}(Embrechts et al.~\cite{em:lw}) 
Consider independent and identically distributed random variables $X_i$, $i=1,\ldots,n$ with common 
cumulative distribution function $F$. 
Assume they are regularly varying with tail index $\b>0$, which means that the right tail $1- F$ of their distribution satisfies
$$
\lim_{x\to\infty} \frac{1- F(ax)}{1- F(x)} =a^{-\b}, \ \text{for all}\ a>0.
$$
Then the risk measure VaR  is \emph{asymptotically subadditive} for $X_1,\ldots,X_n$ if and only if $\b\ge 1$:
$$
\lim_{\alpha\nearrow 1} \frac{VaR_\alpha\big(\sum_{i=1}^n X_i \big)}{\sum_{i=1}^n VaR_\alpha(X_i)} ~\le 1   \quad \Leftrightarrow \quad \b\ge 1.
$$
\end{prop}

\textbf{Ad (ii).} Another important class of distributions which implies the subadditivity of VaR is the class of elliptical distributions.
\begin{prop}(Embrechts et al.~\cite{embrechts2002correlation})\\
Let $X=(X_1,\ldots,X_n)$  be a random vector having an elliptical distribution.
Consider the set of linear portfolios $M=\{Z=\sum_{i=1}^n \lambda_i\, X_i\,|\,\sum_{i=1}^n \lambda_i=1\}$. \\
Then VaR at level $\alpha$ is subadditive on $M$ if\/ $0.5<\alpha <1$ :
$$
VaR_\alpha(Z_1+Z_2)\ \le\ VaR_\alpha(Z_1)+VaR_\alpha(Z_2), \quad Z_1,\, Z_2 \in M.
$$
\end{prop}

\textbf{Ad (iii).} Furthermore, there exists an analogous result for another type of dependence, the Archi\-medean survival copula:
\begin{prop}(Embrechts et al.~\cite{emnw})
Consider random variables $X_i$, $i=1,\ldots,n$ which have the same continuous marginal distribution function $F$. Assume the tail distribution $\bar F=1-F$ is regularly varying with tail index $-\b<0,$ i.e. $\bar F(x)=x^{-\b}G(x)$ for some function $G$ slowly varying at infinity, and assume $(-X_1\ldots,-X_n)$ has an Archimedean copula with generator $\Psi$, which is regulary varying at 0 with index $-\alpha<0.$ Then for all $\alpha>0$, we have
\vspace{-1ex}
\begin{itemize}
\item  VaR is asymptotically subadditive for all $\b>1$;
\vspace{-1ex}
\item  VaR is asymptotically superadditive for all $\b<1$.
\end{itemize}
\end{prop}
\vspace{-1ex}
\noindent Recently, numerical and analytical techniques have been developed in order to evaluate the risk measures VaR and ES under different dependence assumptions regarding the loss random variables.  Such techniques certainly help for a better understanding of  the aggregation and diversification properties of risk measures, in particular  of non-coherent measures such as VaR. 
 In this paper, we do not review all these techniques and results  but refer to Embrechts et al.~\cite{em:pr} and the references therein for an overview.

Nevertheless, it is worth mentioning two recent studies, a new numerical algorithm introduced by Embrechts and co-authors \cite{em:pr} to provide bounds of VaR of aggregated risks, and a study by Kratz \cite{kr:finma}, \cite{kr:normex} on the evaluation of VaR of aggregated heavy tailed risks.
The numerical algorithm introduced in Embrechts et al.~\cite{em:pr} allows for the computation of reliable lower and upper bounds for the VaR of high-dimensional (inhomogeneous) portfolios, whatever the dependence structure is. Quoting the authors, ``surprisingly, additional positive dependence information (like positive correlation) does typically not improve the upper bound substantially. In contrast higher order marginal information on the model, when available, may lead to strongly improved bounds. It is a good news since, in practice, typically  only the marginal loss distribution functions are known or statistically estimated, while the dependence structure between the losses is either completely or partially unknown.'' In Kratz \cite{kr:normex}, a new approach, called Normex, is developed to provide accurate estimates of high quantiles for aggregated independent heavy tailed risks. This method depends only weakly upon the sample size and gives good results for any non-negative tail index of the risks.


\subsection{Robustness}
\label{se:DiscussionRobust}

With respect to the weak topology most of the common risk measures are discontinuous. Therefore and due to some convenient scaling properties detailed in  Proposition 2.1  of Stahl et al. 
\cite{stahl}, in risk management one usually considers robustness as continuity with respect to the Wasserstein distance as defined
by \eqref{eq:wasserstein}.  According to Stahl et al.~\cite{stahl}, variance, Expected Shortfall, Expectiles, and mean are discontinuous with respect to the weak topology whereas VaR at the level $\alpha$ is robust at $F_0$ if $F_0^{-1}$ is continuous at $\alpha.$ Stahl et al. observe that mean, VaR, and Expected Shortfall are continuous with respect to 
the Wasserstein distance and Bellini et al.~\cite{bellini} show that Expectiles are Lipschitz-continuous with respect to the Wasserstein distance with constant $K=\max\{\frac{\alpha}{1-\alpha};\frac{1-\alpha}{\alpha}\}$, which implies continuity with respect to the Wasserstein distance.

With regard to robustness in the sense given in Cont et al.~\cite{cont} (as mentioned in section~\ref{se:robustness1}), Cont et al. demonstrate that historical Expected Shortfall is much more sensitive to the addition of a data point than VaR. Moreover, in contrast to VaR, ES is sensitive to the data point's size. The authors also investigate the impact of the estimation method on the sensitivity and find that historical Expected Shortfall at 99$\%$ level is much more sensitive than  Gaussian and Laplace Expected Shortfall. 
Moreover, they discuss a potential conflict between the requirements of subadditivity, and therefore also coherence, and robustness of a risk measure estimate. 

Taking into account that VaR  because of its definition as a quantile is insensitive to
the sizes of data points that do not fall into a neighborhood of VaR, the observations by Cont et al.\ are not too surprising. 
The notion of ES\footnote{%
Recently, Jadhav et al.~\cite{Jadhav&al} suggested that 'modified expected shortfall' be a robust and coherent variation of ES. 
However, their proof of coherence is wrong. Moreover, Jadhav et al.\ seem to have overlooked that 
Cont et al.\ (\cite{cont}, Section 3.2.3) had looked at modified expected shortfall before and observed that
it is not coherent.%
} was introduced precisely as a remedy to the lack of risk sensitivity of VaR.

Finally, note  that in practice, the estimation of ES will often be based on larger subsamples than the estimation of VaR.
For instance, when using $100,000$ simulation iterations, ES at $99\%$ level is estimated with 1,000 points while the  VaR estimate is based on a small neighborhood of
the 99,000th order statistic.
Moreover,  when investigating empirically the scaling properties of VaR and ES of aggregated financial returns, Hauksson et al.~\cite{hauk:ddms} noticed that
the numerical stability of the scaling exponent was much higher with ES. This observation, in a way, counters the comments of Cont with regard to the amount of data 
needed for estimation. For often one can use high frequency data to precisely estimate ES and then use the scaling property to determine ES for aggregated risks.

\subsection{Elicitability and Conditional Elicitability} \label{se:ESelicitable}


The lack of coherence of Value-at-Risk (VaR), which is up to now the most popular risk measure in practice, draws the attention to another downside risk measure, Expected Shortfall (ES) as defined in \eqref{ES}. 
Expected Shortfall is a coherent risk measure and, in contrast to Value-at-Risk, is sensitive to the severity of losses beyond Value-at-Risk. Nevertheless, as soon as it comes to forecasting and backtesting Expected Shortfall, a potential deficiency arises compared to Value-at-Risk. Gneiting \cite{gneiting} showed that Expected Shortfall is not elicitable. He proved that the existence of convex level sets is a necessary condition for the elicitability of a risk measure and disproved the existence of convex level sets for the Expected Shortfall. 
It is interesting to note that other important risk measures like the variance are not elicitable either (Lambert et al.~\cite{lambert}).
\begin{lem}\label{EScondelic}
For continuous distributions with finite means, ES is conditionally elicitable.
\end{lem}
\textsc{Proof of lemma~\ref{EScondelic}.} \\
Fix $\alpha \in (0,1)$. Let $\mathcal{P} = \{\text{continuous distributions on }\mathbb{R}\text{ with finite means}\}$ and 
$\mathcal{D} = \bigl\{(P,c) \in \mathcal{P} \times \mathbb{R}: P\bigl([c,\infty)\bigr) > 0\bigr\}$.
For continuous distributions $P$, ES simplifies to $\mathrm{ES}_{\alpha}(L)=\mathrm{E}[L|L\geq q_{\alpha}(L)]$ 
where $L$ denotes a generic random variable with
distribution $P$. 
Hence we can rewrite $\mathrm{ES}_{\alpha}(L)$ using $\gamma: \mathcal{D} \to \mathbb{R}$ defined by 
\begin{equation*}
(P,c) \mapsto  \gamma(P,c):=E_P[L|L\ge c]
\end{equation*} 
and $\tilde\gamma: {\cal P} \rightarrow  \R$ defined by
\begin{equation*}
P \mapsto  \tilde\gamma(P):=q_\alpha(L).
\end{equation*}
Since we have $P(L \ge q_\alpha(L)) = 1-\alpha > 0$ for continuous distributions $P$, properties (ii) and (iv) of Definition ~\ref{cond-elicit} are satisfied. Property (i)
holds because quantiles of distributions with finite means are elicitable with strictly consistent scoring function $s(x,y)=(\mathbf{1}_{\{x\geq y\}}-\alpha)(x-y)$ 
(Newey and Powell \cite{neweyP87}).
For fixed $c \in \R$ and $\mathcal{P}_c$ defined as in Definition~\ref{cond-elicit}, an application of Theorem~7 of Gneiting \cite{gneiting} shows that
$\gamma_c$ is elicitable with strictly consistent scoring function
\begin{align*}
s(x,y) & = \bigl(\phi(y) - \phi(x) - \phi'(x)\,(y-x)\bigr)\,\mathbf{1}_{[c,\infty)}(x), \quad\text{where}\\
\phi(x) & = \frac{x^2}{1+|x|}.
	\end{align*} 
This proves property (iii) of Definition~\ref{cond-elicit}.\hfill $\Box$

In practice, Lemma~\ref{EScondelic} implies that, due to its conditional elicitability, 
we can try and forecast ES in a two-step-procedure.
\begin{enumerate}
\vspace{-1ex}
\item
We forecast the quantile as 
$$\hat q_\alpha(L)=\arg\min_x E_P((\mathbf{1}_{\{x\geq L\}}-\alpha)(x-L))$$ using the strictly consistent scoring function $s(x,y)=(\mathbf{1}_{\{x\geq y\}}-\alpha)(x-y)$ from Example~\ref{scoringExples}.
\vspace{-1ex}
\item Taking this result as a fixed value $\hat q_\alpha$, we observe that $E[L|L\ge \hat q_\alpha]$ is just an
expected value. Thus we can use strictly consistent scoring function to forecast $\mathrm{ES}_{\alpha}(L)\approx E[L|L\ge \hat q_\alpha]$. 
If $L$ is square-integrable, the score function simply can be chosen as the squared error such that  
$\mathrm{ES}_{\alpha}(L)\approx \arg\min_x E_{\tilde P}((x-L)^2),$ where $\tilde P(A)=P(A|L\ge \hat q_\alpha).$
\end{enumerate}
The result of this procedure is then a component-wise optimal forecast for the ES.

\begin{lem}\label{varcondelic}
For distributions with finite second moments, the variance is conditionally elicitable.
\end{lem}
\textsc{Proof of lemma~\ref{varcondelic}.} \\
Let $\mathcal{P} = \{\text{Distributions on }\mathbb{R}\text{ with finite second moments}\}$. Defining $\gamma_c$ by
\begin{equation*}
\gamma_c: {\cal P} \rightarrow  \R,\,
P \mapsto  \gamma(P,c):=E_P[(L-c)^2]
\end{equation*} 
and $\tilde\gamma$ by
\begin{equation*}
\tilde\gamma: {\cal P} \rightarrow  \R,\,
P \mapsto  \tilde\gamma(P):=E_P(L)
\end{equation*} 
we can rewrite the variance $E_P((L-E_P(L))^2)$ as $\mathrm{var}(L)=\gamma(P,\tilde\gamma(P))$.
Then $\tilde\gamma$ is elicitable according to Newey and Powell \cite{neweyP87}. For fixed $c$, $\gamma_c$ is elicitable according to Theorem 8~(a) of Gneiting (\cite{gneiting}).
It follows that the variance is conditionally elicitable in the sense of Definition~\ref{cond-elicit}.
\hfill $\Box$
\begin{lem}(see Gneiting \cite{gneiting})
For distributions with finite means, Expectiles are elicitable. 
\end{lem}
As a consequence, by Remark \ref{rm:every}, Expectiles are conditionally elicitable.

\section{Capital allocation and diversification benefits }
\label{se:impact}

\n For risk management purposes, it is useful to decompose the portfolio-wide risk into components \emph{(risk contributions)}
that are associated with the sub-portfolios or assets the portfolio comprises of. There are quite a few approaches
to this problem. See Tasche \cite{tasche3} for an overview. In the following, we discuss the so-called \emph{Euler allocation}
in more detail, as well as the quantification and comparison of the portfolio diversification.

\subsection{Capital allocation using Expected Shortfall or Expectiles}
\label{se:allocation}

\n Tasche \cite{tasche1} argues that from an economic perspective, with a view on portfolio optimization,
it makes most sense to determine risk contributions as sensitivities (partial derivatives). What makes the
definition of risk contributions by partial derivatives even more attractive is the fact that by Euler's theorem
(see Tasche \cite{tasche1} for a statement of the theorem in a risk management context)
such risk contributions add up to the portfolio-wide risk if the risk measure under consideration is homogeneous. 
Technically speaking, we suggest the following definition of risk contributions.

\begin{defi}\label{def:risk.contribution} 
Let $L, L_1, \ldots, L_m$ be random variables such that $L = \sum_{i=1}^m L_i$ and let $\rho$ be a risk measure.
If the derivative $\frac{d \rho(L + h\,L_i)}{d\,h}$ exists for $h=0$ then the risk contribution
	of $L_i$ to $\rho(L)$ is defined by
	\begin{equation}\label{eq:risk.contribution}
		\rho(L_i\,|\,L) \ = \ \frac{d \rho(L + h\,L_i)}{d\,h}\bigg|_{h=0}.
	\end{equation}
\end{defi}
If the derivatives on the right-hand side of \eqref{eq:risk.contribution} all exist for $i=1, 
\ldots, m$ and the
risk measure $\rho$ is homogeneous in the sense of Definition~\ref{coherence} then Euler's theorem implies
\begin{equation*}\label{eq:exhaustive}
	\rho(L) = \sum_{i=1}^m \rho(L_i\,|\,L).
\end{equation*}
Tasche \cite{tasche1} shows that if one of the $L_i$ has a smooth density conditional on the realizations of
the other $L_i$'s then the risk contributions of Expected Shortfall in the sense of Definition~\ref{def:risk.contribution} 
all exist and have an intuitive shape. However, the process of identifying sufficient conditions for 
the existence of partial derivatives of a risk measure and their calculation can be tedious. For coherent
risk measures, Delbaen in \cite{delbaen} advised an elegant method
to determine the risk contributions. In the following theorem we describe the risk contributions to Expected Shortfall. In
Theorem~\ref{th:expectile.contribution} we then use Delbaen's method to derive the risk contributions to
Expectiles.

\begin{theo}(Tasche \cite{tasche1}, Delbaen \cite{delbaen})\label{th:ES.contributions}
If the partial derivative as described in \eqref{eq:risk.contribution} exists for $\rho$ chosen as Expected Shortfall, 
then the risk contribution of a position $L_i$ to the portfolio's Expected Shortfall can be calculated as
\begin{equation*}
\mathrm{ES}_{\alpha}(L_i|L) =\mathrm{E}[L_i|L\geq q_\alpha(L)]
\end{equation*}
\end{theo}

With Delbaen's approach, we can also derive the capital allocation for Expectiles. See Martin \cite{Martin} for an alternative approach based 
on saddlepoint approximation.
\begin{theo}\label{th:expectile.contribution}
If the partial derivative as described in \eqref{eq:risk.contribution} exists for $\rho = e_{\tau}$, then, for $1/2 \le \tau < 1$, the risk contribution of a position $L_i$ to the portfolio's Expectile can be calculated as
\begin{equation}\label{rcontribexpectile}
e_{\tau}(L_i|L)=\dfrac{\tau\mathrm{E}[L_i\mathbf{1}_{\{L > e_{\tau}(L)\}}]+
(1-\tau)\mathrm{E}[L_i\mathbf{1}_{\{L\le e_{\tau}(L)\}}]}{\tau P[L > e_{\tau}(L)]+(1-\tau)P[L\le e_{\tau}(L)]}.
\end{equation}
\end{theo}

\textsc{Proof of Theorem \ref{th:expectile.contribution}.}\\
The proof follows Delbaen's method (Delbaen \cite{delbaen}). Recall that the weak subgradient of a convex function $f:L^\infty(\Omega)\rightarrow \R$ at $X\in L^\infty(\Omega)$ (see Section~8.1 of Delbaen \cite{delbaen}), is defined as:
\begin{multline*}
	\nabla f(X) = \{\varphi: \varphi\in L^1(\Omega)\text{\ such that for all} \ Y\in L^\infty(\Omega), \  f(X+Y) \ge f(X) + \mathrm{E}[\varphi\,Y]\}.
\end{multline*}
In order to identify the subgradient of the risk measure $e_{\tau}$, we note that 
\begin{itemize}
\vspace{-2ex}
	\item $e_{\tau}$ is a law-invariant coherent risk measure,
	\vspace{-1ex}
	\item as shown in Jouini et al.~\cite{jouini2006law}, $e_{\tau}$ has the so-called Fatou-property,
	\vspace{-1ex}
	\item as shown in Bellini et al.~\cite{bellini}, we have  that
	\begin{equation*}
	\begin{split}
		e_{\tau}(L) & = \max\bigl\{\mathrm{E}[\varphi\,L]: \varphi \in M_\tau\bigr\}, \ \text{with}\\
		M_\tau & = \bigl\{\varphi \ge 0 \text{\ is bounded with}\ \mathrm{E}[\varphi]=1
		\text{\ and}\ \tfrac{\sup \varphi}{\inf \varphi}\le \max\bigl(\tfrac{\tau}{1-\tau}, \tfrac{1-\tau}{\tau}\bigr)\bigr\},
	\end{split}
	\end{equation*}
	\vspace{-2ex}
	\item as shown in Bellini et al.~\cite{bellini}, for $\displaystyle \bar{\varphi} = \frac{\tau\,
		\mathbf{1}_{\{L>e_\tau(L)\}}+(1-\tau)\,\mathbf{1}_{\{L\le e_\tau(L)\}}}%
			{\tau\,\mathrm{P}[L>e_\tau(L)]+(1-\tau)\,\mathrm{P}[L\le e_\tau(L)]}$, we have
				$\bar{\varphi} \in M_\tau$ and $e_\tau(L) = \mathrm{E}[\bar{\varphi}\,L]$.			
\end{itemize}
\vspace{-1ex}
Theorem 17 of Delbaen \cite{delbaen} now implies that $\bar{\varphi}$ is an element of $\nabla e_\tau(L)$, i.e.\ it holds for
all bounded random variables $L^\ast$ that 
\begin{equation*}
	e_\tau(L+L^\ast) \ \ge \ e_\tau(L) + \mathrm{E}[\bar{\varphi}\,L^\ast].
\end{equation*}
From Proposition~5 of Delbaen \cite{delbaen} it follows that, if $\nabla e_\tau(L)$ has only one element, then we have
\begin{equation}\label{eq:deriv}
	\frac{d \,e_\tau(L+h\,L^\ast)}{d\,h}\bigg|_{h=0} \ = \ \mathrm{E}[\bar{\varphi}\,L^\ast].
\end{equation}
Taking $L^\ast = L_i$ in equation \eqref{eq:deriv} implies \eqref{rcontribexpectile}. \hfill $\Box$

The proof of Theorem \ref{th:expectile.contribution} shows that risk contributions for Expectiles (and also for Expected Shortfall) can still be defined,
even if the derivatives in the sense of Definition~\ref{def:risk.contribution}
do not exist. This may happen if the distribution of the loss variable is not smooth (e.g.\ not continuous). Then
the subgradient set $\nabla e_\tau(L)$ may contain more than one element such that there is no unique candidate vector for 
the risk contributions. See Kalkbrener \cite{kalkbrener} for more details on this approach to risk contributions for coherent risk measures.

\subsection{Diversification benefits}

In risk management, evaluating  diversification benefits properly  is key to both insurance and investments, since risk diversification may reduce a company's need for risk-based capital. 
To quantify and compare the diversification of portfolios, indices have been defined, such as  the closely related notions of 
{\it diversification benefit} defined by B\"urgi et al.~\cite{BuergiEtAl}, and the  \emph{diversification index} by Tasche \cite{tasche3}. 
Both indices are not universal risk measures and depend on the choice of the risk measure  and on the number of the underlying risks in the portfolio.

As mentioned earlier, subadditivity and comonotonic additivity of a risk measure are important conditions for proper representation of diversification effects. 
In this case, capital allocation as introduced in Section~\ref{se:allocation} can be helpful for identifying risk concentrations.

\n Let us define the diversification index (Tasche \cite{tasche3}):
\begin{subequations}
\begin{defi}
Let $L_1,\ldots, L_n$ be real-valued random variables and let $L=\sum_{i=1}^n L_i$. If $\rho$ is a risk measure such that $\rho(L), \rho(L_1),\ldots,\rho(L_n)$ are defined, then 
\begin{equation*}
DI_{\rho}(L)=\dfrac{\rho(L)}{\sum_{i=1}^n\rho(L_i)}
\end{equation*}
denotes the diversification index of portfolio L with respect to the risk measure $\rho$.\\ 
If risk contributions $\rho(L_i|L)$  of $L_i$ to $\rho(L)$ (see Definition~\ref{def:risk.contribution}) exist, then 
\begin{equation*}
DI_{\rho}(L_i|L)=\dfrac{\rho(L_i|L)}{\rho(L_i)}
\end{equation*}
denotes the marginal diversification index of subportfolio $L_i$ with respect to the risk measure $\rho.$
\end{defi}
\end{subequations}
For the case of a homogeneous, subadditive, and comonotonically additive risk measure, Tasche derived the following properties of the diversification index:
\begin{propties}(Tasche \cite{tasche3})\label{indexD} 
Let $\rho$ be a homogeneous, subadditive, and comonotonically additive risk measure. Then
\begin{itemize}
\item $DI_{\rho}(L)\leq 1$ (due to subadditivity).
\item $DI_{\rho}(L)\approx 1$ indicates that $L_1, \ldots,L_n$ are `almost' comonotonic. The closer to one the index of diversification is, 
the less diversified is the portfolio.
\item If $DI_{\rho}(L_i|L)<DI_{\rho}(L)$, then there exists $\epsilon_i>0$ such that  $DI_{\rho}(L+hL_i)<DI_{\rho}(L)$, for all $0<h<\epsilon_i$.
\end{itemize}
\end{propties}
It is not clear how far below 100\% the diversification index should be to indicate high diversification because, in the presence of
undiversifiable risk, even a large optimised portfolio might still have a relatively high index. 
Nonetheless, comparison between marginal diversification indices and the portfolio's diversification index can be useful to detect unrealized diversification potential. Hence, instead of investigating the absolute diversification index, it might be better to look for high unrealized diversification potential as a criterion to judge a portfolio as highly concentrated.

Note that risk measures like standard deviation or Expectiles would show a 100\% diversification index for portfolios with perfectly
linearly correlated positions but not for comonotonic positions with less than perfect linear correlation. Hence, for risk measures that are not 
comonotonically additive there is a danger of underestimating lack of diversification due to non-linear dependence. 

 A notion similar to the diversification index was proposed in B\"urgi et al.~\cite{BuergiEtAl} to quantify the diversification performance of a 
portfolio of risks.  B\"urgi et al.\ define the  notion of diversification benefit, denoted by $D\!B$, of a portfolio $L=\sum_{i=1}^n L_i$ as
\begin{equation*}\label{eq:DBn}
D\!B(L)=1-\frac{RAC_\rho (\sum_{i=1}^{n}{L_{i}})}{\sum_{i=1}^{n}{RAC_\rho (L_{i})}}
\end{equation*}
where $RAC$ denotes the Risk Adjusted Capital defined as the least amount of additional capital needed to prevent 
a company's insolvency  at a given level of default probability:  
\begin{equation*}
RAC_\rho(L)=\rho(L)-E(L)
\end{equation*}
where $\rho(L)$ denotes the risk measure chosen for $L$.
Clearly, $D\!B$ has properties very similar to the properties of the the diversification index, namely:
\begin{propties}(B\"urgi et al.~\cite{BuergiEtAl})\label{Db}  
Let $\rho$ be a homogeneous, subadditive, and comonotonically additive risk measure. Then
\begin{itemize}
\item $0\le D\!B(L)\leq 1$ (due to subadditivity)
\item The interpretation of the diversification benefit is straightforward, namely
\begin{equation*}
	D\!B(L)=
\left\{\begin{array}{cl}
1 & \text{indicates full hedging}\\
0 & \text{indicates comonotonic risks}\\
x\in ]0,1[ &  \text{indicates that there is  $100\,x\%$ of capital reduction}\\
&\text{due to diversification.} 
\end{array}
\right.
\end{equation*}
Hence the higher $D\!B(L)$, the higher the diversification (in contrast to the diversification index $DI_{\rho}$).
\end{itemize}
\end{propties}

The same comments apply to both Properties~\ref{indexD} and Properties~\ref{Db}. Both indices  depend not only on the choice of $\rho$ and on the portfolio size $n$, but even more strongly on the dependence structure between the risks. 
Neglecting dependence may lead to a gross underestimation of RAC. 
 This has been analytically illustrated  with a simple model in Busse et al.~\cite{BusseEtAl}, where 
it is demonstrated that introducing dependence between the risks drastically reduces  the diversification benefits.  
 
 When it comes to comparing the consequences of choosing VaR and ES respectively for the measurement of diversification benefits, 
we can really see the limitation of VaR as a risk measure. Even if there is a part of the risk that is undiversifiable, 
VaR might not catch it as demonstrated in Proposition~3.3 of Emmer and Tasche \cite{emmer&tasche}. In Busse et al.~\cite{BusseEtAl}, VaR shows a diversification benefit for a very high number $n$ of risks, 
while ES does not decrease for this range of $n$, thus correctly reflecting the fact that the risk cannot completely be diversified away.
 
Moreover, the type of dependence does matter. Linear dependence (measured with the linear correlation) 
cannot accurately describe  dependence between extreme risks, in particular in times of stress.  Neglecting the 
non-linearity of dependence may lead to an overestimation of the diversification benefits. This is well described by B\"urgi et 
al.~\cite{BuergiEtAl} who consider elliptical and Archimedean copulae  for risk modelling and compare their impacts 
on the evaluation of RAC and hence also on the diversification benefit.

\section{Backtesting: which methods can be used?} \label{se:backtesting}

What does backtesting mean? According to Jorion \cite{jorion}, it is a set of statistical procedures designed to 
check if the realized losses, observed ex post, are in line with VaR forecasts. We may of course extend this definition to any risk measure.

Recently, Gneiting \cite{gneiting} has raised a potential issue with direct backtesting when using 
Expected Shortfall (ES) as a risk measure. This is not an issue for risk measures like VaR or Expectiles  because of their elicitability, as seen previously. Is it a real issue in practice for ES? 
On the one hand, Acerbi \& Sz\'ekely \cite{ac:sz} recently have argued that actually elicitability (or lack of elicitability)
is not relevant for backtesting of risk measures but rather for comparing the forecast performance of different estimation methods. On the other hand, 
some financial institutions, 
in particular reinsurance companies, have addressed the problem of backtesting ES by using probability distribution forecasts for checking the output of their internal models. 
%
%
Nevertheless, if one still wants to stick to point forecasts only for ES, we propose an empirical approach that consists in approximating ES with quantiles  -- see Section \ref{se:backVaR}.

Furthermore, as observed in Section~\ref{se:ESelicitable}, ES is a combination of two elicitable components, since it is conditionally elicitable. A natural approach to the backtesting of 
ES therefore is to use the algorithm described in Section~\ref{se:ESelicitable}, where we backtest both components separately according to their associated respective scoring functions. Here as the first component we backtest the quantile. Then, taking the result for the quantile as a fixed value, we can backtest  the ES, since it is then just a mean, which has the quadratic error as strictly consistent scoring function.
 
More generally, the choice of the backtesting method should depend on the type of forecast. There are backtesting methods for:
\begin{itemize}
\vspace{-1ex}
\item[(i)] {\it Point forecasts} for the value of a variable; they are 
usually represented as the conditional expectation $\E[Y_{t+k}~|~ {\cal F}(Y_s,s\le t)]$ where ${\cal F}(Y_s,s\le t)$ represents the available information up to time $t$ on the time series $Y$. 
There is a huge amount of literature, notably in econometrics,  on point forecasts and on well-established methods for their out-of-sample backtesting 
(e.g. Clements and Hendry \cite{cl:he} or Elliott et al.~\cite{eg:tim}). 
\vspace{-1ex}
\item[(ii)]  {\it Probability range forecasts} or {\it interval forecasts} (e.g.\ forecasts of Value-at-Risk or of Expected Shortfall); 
they project an interval in which the forecast value is expected to lie with some probability $p$  (e.g. the interval $(-\infty, VaR_p(Y_{t+k})]$ where $VaR_p(Y_{t+k})$ is the projected $p$-quantile of $Y_{t+k}$).
Much work, in particular with regard to backtesting, has been done on interval forecasts in the last 15 years. A good reference on this topic is Christoffersen (\cite{christof}). Backtesting for VaR has been well developed, due to the interest of the financial industry in this risk measure. We refer e.g. to Dav\'e and Stahl \cite{dave}, and, for a review on backtesting procedures for VaR, to Campbell \cite{campbell}. 
\vspace{-1ex}
\item[(iii)] {\it Forecasts of the complete probability distribution} $\P[Y_{t+k}\le . ~|~ {\cal F}(Y_s,s\le t)]$ or its probability density function, if existing.
\end{itemize}
\vspace{-1ex}

It is worth noticing that if there is a solution to (iii) then there are also solutions for (i) and (ii), and that (iii) makes it possible to backtest ES, avoiding then the issue raised by Gneiting (\cite{gneiting}) for the direct backtesting of ES. 

In contrast to VaR, ES is sensitive to the severity of losses exceeding the threshold VaR because the risk measure ES corresponds to the full tail of a distribution. Hence, seen as a part of the distribution beyond a threshold, the accuracy of the forecast of ES may be directly checked using tests on the accuracy of forecasts of probability distributions
(see Tay and Wallis \cite{TayWallis} and Gneiting and Katzfuss \cite{gneitingK} for general discussions of this approach). Note that the tail of the distribution might be evaluated through a Generalized Pareto Distribution (GPD) above a high threshold via the Pickands theorem (see Pickands \cite{pickands} or Embrechts et al.~\cite{em:km}).

In the following, we provide more detail on (ii) and (iii).

\subsection{Backtesting VaR and ES}
\label{se:backVaR}

\paragraph{Backtesting VaR.}

%

As mentioned in Example~\ref{scoringExples}, VaR is elicited by the weighted absolute error scoring function (see 
Thomson \cite{thomson}, Saerens \cite{saerens}, or Gneiting \cite{gneiting} 
for details), 
characterizing VaR as an optimal point forecast. This allows for the comparison of different forecast methods. However,
in practice, we have to compare VaR predictions by a single method with observed values to assess the quality of the predictions. 

A popular procedure is based on the so-called violation process briefly described here.
Since by definition of VaR, assuming a continuous loss distribution, we have
$\displaystyle \P(L> VaR_\alpha(L)) =1-\alpha$, it follows that  the probability of a  violation of VaR is $1-\alpha$.
We define the \emph{violation process} of VaR as
\begin{equation*}
	I_t(\alpha)=\mathbf{1}_{\big\{L(t)> VaR_\alpha(L(t))\big\}}.
\end{equation*}
Christoffersen (\cite{christof}) showed that VaR forecasts are valid if and only if the violation process $I_t(\alpha)$ satisfies two conditions:
\begin{itemize}
\vspace{-2ex}
	\item the unconditional coverage hypothesis: $\E[I_t(\alpha)]=1-\alpha$, and
	\vspace{-1ex}
	\item the independence condition: $I_t(\alpha)$ and $I_s(\alpha)$ are independent for $s\ne t$
\end{itemize}
\vspace{-2ex}
  Under these two conditions,  the $I_t(\alpha)$'s are independent and identically distributed  Bernoulli random variables with success probability $1-\alpha$. Hence the number of violations has a Binomial distribution.

This means in practice to consider an estimate of the violation process by replacing VaR by its estimates and check that this process behaves like 
independent and identically distributed Bernoulli random variables with violation (success) probability close to $1-\alpha$. If the proportion of  VaR violations is not significantly  
different from $1-\alpha$, then we conclude that the estimation/prediction method is reasonable.

However, the above independence condition might be violated in practice, such that the general way of computing VaR as an unconditional quantile from the historical sample
seems questionable. That is why various tests on the independence assumption have been proposed in the literature, as e.g.~one developed by Christoffersen and Pelletier 
(see \cite{christofPelletier}), based on the duration of days between the violations of the VaR thresholds.

\paragraph{Backtesting ES.} 

A similarly simple approximative approach to the backtesting of ES might be based  on a representation of ES as integrated VaR (Acerbi and Tasche \cite{ac:ta}, Proposition~3.2):
\begin{eqnarray} \label{ESapprox}
\mathrm{ES}_\alpha(L) & = & \frac1{1-\alpha}\int_\alpha^1 q_u(L)\,du \notag\\
& \approx & \frac14 \,\left[\, q_\alpha(L)+q_{0.75\,\alpha+0.25}(L) +  q_{0.5\,\alpha+0.5}(L)+q_{0.25\,\alpha+0.75}(L)\,\right],
\end{eqnarray}
where $q_\alpha(L)= VaR_\alpha(L)$.
Hence, if $q_\alpha(L)$, $q_{0.75\,\alpha+0.25}(L)$, $q_{0.5\,\alpha+0.5}(L)$, and $q_{0.25\,\alpha+0.75}(L)$ are successfully backtested, then also the estimate of $\mathrm{ES}_\alpha(L)$ can be considered reliable subject to a careful manual inspection of the observations in the upper 0.25\% tail of the observed sample. The upper tail observations must anyway be manually inspected in order to separate data outliers from genuine far tail observations. In so far, the suggested procedure provides a reasonable combination
of statistical testing and human oversight. Compared to the test procedures suggested in Acerbi and Sz\'ekely \cite{ac:sz}, it has the advantage of not relying on Monte-Carlo simulation
for the statistical test. 

Do four supporting points suffice in the linear approximation to ES by different VaRs in \eqref{ESapprox}? Actually, 
the power of the joint test for VaR violations on the supporting points will decline with the number of supporting points chosen but increase with the size of the sample of 
available observations. Hence, the number of supporting points must be determined on a case by case basis with a view on the sample size.

The approach based on \eqref{ESapprox} is attractive not only for its simplicity but also because 
it illustrates the fact that for the same level of certainty a much longer sample is needed for the validation of  
$\mathrm{ES}_\alpha(L)$ than for $\mathrm{VaR}_\alpha(L)$ (see also Yamai and Yoshiba \cite{Yamai&Yoshiba}).
The Basel Committee suggests a variant of this ES-backtesting approach which is based on testing level violations for two quantiles 
at 97.5\% and 99\% level \cite{basel2013fundamental}.

\subsection{Backtesting distribution forecasts}

Let us outline a method for the out-of-sample validation of distribution forecasts, based on the L\'evy-Rosenblatt transform, named also 
\emph{Probability Integral Transform (PIT).} 
As pointed out before, this methodology is important since testing the distribution forecasts could be helpful, in particular for tail-based risk measures like ES. 

The use of the PIT for backtesting financial models is relatively recent. The foundations were laid by  Diebold and coauthors.
Diebold et al.~\cite{diebold2} tackled the problem of density forecast evaluation from a risk management perspective, 
suggesting a method for testing continuous distribution forecasts in finance, based on the uniform distribution of  the L\' evy-Rosenblatt transform (or PIT) 
(L\' evy \cite{levy} and Rosenblatt \cite{rosen}). Applying the L\' evy theorem to the PIT, they observed that if  a sequence of distribution forecasts coincides with the sequence of unknown conditional laws 
that have generated the observations, then the sequence of PIT are independent and identically distributed ${\cal U} (0,1)$.
In Diebold et al.~\cite{diebold3}, they extended the density forecast evaluation to the multivariate case, involving cross-variable interactions 
such as time-varying conditional correlations, and provided conditions under which a technique of density forecast `calibration' 
can be used to improve deficient density forecasts. They finally applied the PIT method on high-frequency financial data (volatility forecasts) to illustrate its application.
Note that the definition of PIT has been generalized for not necessarily continuous cumulative distribution functions (cdf) (see Gneiting and Ranjan \cite{gneitingRanjan13} and references therein).

Nevertheless, there was still some gap to fill up before a full implementation and use in practice. Blum in his PhD thesis \cite{blum} studied various issues left open, and proposed and validated mathematically a method based on PIT also in situations with overlapping forecast intervals and multiple forecast horizons. Blum illustrated this in his thesis dealing with economic scenario generators (ESG).
Typically, financial institutions make use of scenario generators, producing thousands of scenarios, each one having its own forecast value for a certain value at a certain future time. 
Recall that the scenarios are constructed by simulating the iid innovations of the underlying process. Those simulated values define an empirical distribution, which represents a distribution forecast. Hence the backtesting will be done on the obtained distribution; it is an out-of-sample backtesting of distribution forecasts.
For details of the methodology, we refer to Blum \cite{blum}, SCOR Switzerland \cite{scor} and the references therein and only summarize the main steps in the following.  

From the values obtained from all the scenarios, we deduce the empirical distribution denoted by $\hat\Phi_i$, which is assumed to converge to the marginal cdf $\displaystyle \Phi_i$ defined by $\displaystyle \Phi_i(x)=\P(X_i \le x ~|~ {\cal F}_{i-m})$ where $X_i$ corresponds to the scenario forecast of a variable $X$ at out-of-sample time point $t_i$ and $\displaystyle {\cal F}_{i-m}$ to the information available up to time $t_{i-m}$ from the simulation start, $m$ being the number of forecast steps. 
Hence, at out-of-sample time point $t_i$, we make use of $\hat\Phi_i$, when identifying the distribution at time $t_i$ as the one computed at the previous time $t_{i-m}$, and a newly observed value $x_i$. 

Now we apply the PIT to build the random variables $Z_i:=\hat\Phi_i(X_i)$, with known realizations $\hat\Phi_i(x_i)$. These have been proved by Diebold et al.~\cite{diebold2}, Diebold et al.~\cite{diebold3} to be independent and identically ${\cal U}(0,1)$-distributed  whenever the conditional distribution forecast  $\Phi_i(.)$ coincides with the true process by which the historical data have been generated.

For practical purposes, it then suffices to test if the PIT-transformed variables $Z_i$ are independent and identically ${\cal U}(0,1)$-distributed. If one of these conditions is rejected, the model does not pass the out-of-sample test. As noted by Diebold and Mariano \cite{diebold1}, this is not a test on the model, so it does not mean the model is valueless. Rejection only means that there may be a structural difference between the in-sample and out-of-sample periods  
or that the model does not hold up to the full predictive data.

Various statistical tests are possible, like standard tests such as the $\chi^2$ test for uniformity or the Kendall-Stuart test for the significance of the autocorrelations. 
Going on with the Diebold et al.\ methodology, their non-parametric test, proposed in Diebold et al.~\cite{diebold2} (see also Diebold et al.~\cite{diebold3} for the multivariate case), 
may also be useful. This test  consists of comparing histograms obtained from $Z_i$ and ${\cal U} (0,1)$ respectively, and of detecting deviations from the independence property when considering correlograms of the $Z_i$ and their lower integer powers.  

Note that tests based on PIT have some limitation due to serial correlation. One way to overcome this issue is for instance, 
as suggested in SCOR Switzerland \cite{scor}, to generate realistic forecast scenarios via refined bootstrapping. 

Many other results have enriched the literature on distribution backtesting (see e.g. Elliott et al.~\cite{eg:tim} , Gneiting and Katzfuss \cite{gneitingK}). We may mention two other methods completing our review, one based on the notion of scoring  (see e.g. Gneiting and Raftery \cite{gneitingR}, Gneiting and Ranjan\cite{gneitingRanjan}, Amisano and Giacomini \cite{am:gia}, or the survey paper Gneiting and Katzfuss \cite{gneitingK}), the other mixing the scoring and PIT approaches (see Gneiting and Katzfuss \cite{gneitingK}). 
We already introduced the concept of scoring function $s$ in Definition~\ref{scoring}. When using it for backtesting purposes,  
we modify it to measure the loss function $s(f, Y)$  whose arguments are the density forecast $f$ and the realization $y$ of the future observation $Y$.

\section{Conclusion}
\label{se:concl}

In this paper, we have listed a number of properties that are commonly considered must-haves for good risk measures: 
coherence, comonotonic additivity, robustness, and elicitability. We have then revisited the popular risk measures Value-at-Risk (VaR) and 
Expected Shortfall (ES)
as well as the recently suggested Expectiles and checked which of these properties they satisfy: 
\begin{itemize}
\item It is well-known that VaR lacks subadditivity in general and, therefore, might fail to appropriately account for risk concentrations. 
However, we found that for many practical applications this might not be a serious issue, as long as the underlying risks have a finite variance, or,  in some cases, a finite mean. 
 The fact that VaR does not cover tail risks `beyond' VaR is a more serious deficiency although ironically it makes VaR a risk measure that is 
more robust than the other risk measures we have considered.
This deficiency can be particularly serious when one faces choices of various risks with different tails. VaR and ES will present different optimal results that are well known to be sub-optimal in terms of risk for VaR (e.g.\ McNeil et al.~\cite{McN:fe}, Example 6.7).
\item ES makes good for the lack of subadditivity and sensitivity for tail risk of VaR but has recently be found to be not elicitable. This means 
that backtesting of ES is less straightforward than backtesting of VaR. 
We have found that nonetheless there are a number of feasible approaches to the backtesting of ES (e.g.~based on distribution forecasts, 
linear approximation of ES with VaR at different confidence levels, or directly with Monte-Carlo tests).
However, it must be conceded that to reach the same level of certainty more validation data is required for  ES than for VaR.
\item Expectiles have been suggested as coherent and elicitable alternatives to ES. However, while Expectiles indeed have a number of attractive 
features, their underlying concept is less intuitive than the concepts for VaR or ES. In addition, Expectiles are not comonotonically additive which 
implies that in applications they may fail to detect risk concentrations due to non-linear dependencies.
\end{itemize}
To conclude, we have found that among the risk measures we discussed, ES seems the best for use in practice, despite some caveats with regard to its estimation and backtesting, 
which can be carefully mitigated. We have not found sufficient evidence to justify an all-inclusive replacement of ES by its recent competitor Expectile. Nonetheless, it is certainly worthwhile to keep in mind Expectiles as alternatives to ES and VaR in specific applications.

{\small {\bf Acknowledgement.} We thank two referees for their careful reading of the manuscript and suggestions that helped improving the presentation of the paper. We are also grateful to Michel Dacorogna for interesting discussions on backtesting in practice. Partial support from RARE-318984 (an FP7 Marie Curie IRSES Fellowship) is kindly acknowledged.}

\end{document}